\documentclass{article}
\usepackage{arxiv}
\usepackage[ruled,linesnumbered,onelanguage]{algorithm2e}
\usepackage{graphicx}
\usepackage{hyperref}
\usepackage{listings}
\usepackage{xcolor}
\usepackage{subcaption}
\usepackage{times}
\usepackage{todonotes}
\usepackage{url} 
\usepackage{multirow}  
\usepackage{colortbl}
\usepackage{wrapfig}
\usepackage{floatflt}

\newcommand{\secref}[1]{Section~\ref{#1}}
\newcommand{\ie}{\emph{i.e.,}\xspace}

\newcommand{\ttt}[1]{\texttt{#1}}

\definecolor{white}{rgb}{1,1,1}
\definecolor{backgroundColour}{rgb}{0.95,0.95,0.92}
\definecolor{mGray}{rgb}{0.5,0.5,0.5}
\definecolor{greencomment}{rgb}{0.247,0.498,0.372}

\lstdefinestyle{JavaStyle}{
    backgroundcolor=\color{backgroundColour}, 
  language=Java,
  tabsize=4,
  breaklines=true,
  breakatwhitespace=true,
  escapechar=|*,
  numbers=left,                    
  numbersep=5pt,
  stepnumber=1,
  basicstyle=\footnotesize\ttfamily,
  numberstyle=\footnotesize\ttfamily\color{mGray},
  aboveskip=\baselineskip,
  captionpos=b,
  frame=single,
  columns=fullflexible,
  showstringspaces=false,
  extendedchars=true,
  breaklines=true,
  showtabs=false,
  showspaces=false,
  identifierstyle=\ttfamily,
  keywordstyle=\color[rgb]{0.498,0.0,0.333},
  stringstyle=\color[rgb]{0.165,0.0,0.999},
  commentstyle=\color[rgb]{0.247,0.498,0.372},
  morekeywords={}%{byteArrayCompare}
}
\begin{document}
\title{BISM: Bytecode-Level Instrumentation for Software Monitoring}
\author{
  Chukri~Soueidi\\
  Univ.~Grenoble~Alpes,~Inria,\\ 
  CNRS,~Grenoble~INP,~LIG\\ 
  38000~Grenoble\\ 
  France \\
  \texttt{chukri.a.soueidi@inria.fr} \\
  \And
  Ali~Kassem\\
  Univ.~Grenoble~Alpes,~Inria,\\ 
  CNRS,~Grenoble~INP,~LIG\\ 
  38000~Grenoble\\ 
  France \\
  \texttt{ali.kassem@inria.fr} \\
  \And
  Yli\`{e}s~Falcone\\
  Univ.~Grenoble~Alpes,~Inria,\\ 
  CNRS,~Grenoble~INP,~LIG\\ 
  38000~Grenoble\\ 
  France \\
  \texttt{ylies.falcone@inria.fr} \\
}
\maketitle              
\begin{abstract}
BISM (Bytecode-Level Instrumentation for Software Monitoring) is a light\-weight bytecode instrumentation tool that features an expressive high-level control-flow-aware instrumentation language. 
The language follows the aspect-oriented programming paradigm by adopting the joinpoint model, advice inlining, and separate instrumentation mechanisms. 
BISM provides joinpoints ranging from bytecode instruction to method execution, access to comprehensive static and dynamic context information, and instrumentation methods. 
BISM runs in two instrumentation modes: build-time and load-time. 
We demonstrate BISM effectiveness using two experiments: a security scenario and a general runtime verification case. 
The results show that BISM instrumentation incurs low runtime and memory overheads. 

\end{abstract}  
\section{Introduction}
Instrumentation is essential to the software monitoring workflow~\cite{BartocciFFR18}.
Instrumentation allows extracting information from a running software to abstract the execution into a trace fed to a monitor.
Depending on the information needed by the monitor, the granularity level of the extracted information may range from coarse (e.g., a function call) to fine (e.g., an assignment to a local variable, a jump in the control flow).

For software instrumentation, aspect-oriented programming (AOP)~\cite{KiczalesLMMLLI97} is a popular and convenient paradigm where instrumentation is a cross-cutting concern.
For Java programs, runtime verification tools~\cite{FalconeKRT18,BartocciFBCDHJK19} have for long relied on AspectJ~\cite{KiczalesHHKPG01}, which is one of the reference AOP implementations for Java.
AspectJ provides a high-level pointcut/advice model for convenient instrumentation. 
However, AspectJ does not offer enough flexibility to perform some instrumentation tasks that require to reach low-level code regions, such as bytecode instructions, local variables of a method, and basic blocks in the control-flow graph (CFG).

Yet, there are several low-level bytecode manipulation frameworks such as ASM~\cite{BrunetonASM02} and BCEL~\cite{BCEL}. %\footnote{\url{https://commons.apache.org}}.
However, writing instrumentation in such frameworks is tedious and requires expertise on the bytecode.  
Other bytecode instrumentation frameworks, from which DiSL~\cite{MarekVZABQ12} is the most remarkable, enable flexible low-level instrumentation and, at the same time, provide a high-level language. 
However, DiSL does not allow inserting bytecode instructions directly but provides custom transformers where a developer needs to revert to low-level bytecode manipulation frameworks. 
This makes various scenarios tedious to implement in DiSL or incur a considerable bytecode overhead.

\paragraph{Contributions.}
In this paper, we introduce BISM (Bytecode-Level  Instrumentation for Software Monitoring), a lightweight bytecode instrumentation tool that features an expressive high-level instrumentation language. 
The language inspires from the AOP paradigm by adopting the joinpoint model, advice inlining, and separate instrumentation mechanisms.  
In particular, BISM provides a separate class to specify instrumentation code, and offers a variety of \emph{joinpoints} ranging from bytecode instruction to basic block and method execution. 
BISM also provides access to a set of comprehensive joinpoints-related \emph{static} and \emph{dynamic contexts} to retrieve some relevant information, and a set of \emph{instrumentation methods} to be called at joinpoints to insert code, invoke methods, and print information.   
BISM is control-flow aware.
That is, it generates CFGs for all methods and offers this information at joinpoints and context objects. 
Moreover, BISM provides a variety of control-flow properties, such as capturing conditional jump branches and retrieving successor and the predecessor basic blocks. 
Such features help instrumenting tools using a control-flow analysis, for instance, in the security domain, to detect control-flow attacks, such as test inversions and arbitrary jumps.

We demonstrate BISM effectiveness using two complementary experiments.
The first experiment shows how BISM can be used to instrument for a security scenario, more particularly, to detect test inversions in the control-flow of AES (Advanced Encryption Standard).  
The second experiment demonstrates a general runtime verification case, where we used BISM to instrument seven applications from DaCapo benchmark~\cite{DaCapo06} to verify the classical HasNext, UnsafeIterator and SafeSyncMap properties. 
We also compare BISM's performance to DiSL using three metrics: size, memory footprint, and runtime of the instrumented code.    
The results show that BISM instrumentation incurs a smaller size and memory footprint.  
Regarding the runtime of the instrumented code, in load-time instrumentation, BISM always performs better, and in build-time instrumentation, BISM performs better except for two out of seven benchmarks in the second experiment. 
We observe, in load-time, that the two tools perform similarly when many classes are in the scope of instrumentation but not affected. 
This stems from (1) DiSL's faster generation of static objects and (2) the static analysis performed by BISM on all classes in the scope, even if not used. 
\paragraph{Paper organization.}
The rest of the paper is organized as follows. 
Section~\ref{sec:bism-design} overviews the design goals and features of BISM.  
Section~\ref{sec:bism-language} presents the language featured by BISM.  
Section~\ref{sec:framework} presents the implementation of BISM. 
Section~\ref{sec:casestudies} presents case studies and a comparison between BISM and DiSL. 
Section~\ref{sec:related-work} discusses related work. 
Finally, Section~\ref{sec:conclusion} concludes. % and outlines avenues for future work. 
\section{BISM Design and Features}
\label{sec:bism-design}

BISM is implemented on top of ASM~\cite{BrunetonASM02} with the following goals and features. 
\paragraph{Instrumentation mechanism.} 
BISM language follows the AOP paradigm. 
It provides a mechanism to write separate instrumentation classes. 
An instrumentation class specifies the instrumentation code to be inserted in the target program at chosen joinpoints. 
BISM offers joinpoints that range from bytecode instruction to basic block and method execution.  
It also offers several instrumentation methods and, additionally, accepts instrumentation code written in the ASM syntax. 
The instrumentation code is eventually compiled by BISM into bytecode instructions and inlined in the target program at the exact joinpoint locations. 

\paragraph*{Access to program context.}
BISM offers access to complete static information about instructions, basic blocks, methods, and classes. 
It also offers dynamic context objects that provide access to values that will only be available at runtime such as values of local variables, stack values, method arguments, and results. 
Moreover, BISM allows accessing instance and static fields of these objects. 
Furthermore, new local variables can be created within the scope of a method to pass values between joinpoints.

\paragraph{Control flow context.}
BISM generates the CFGs of target methods out-of-the-box and offers this information within joinpoints and context objects. 
In addition to basic block entry and exit joinpoints, BISM provides specific control-flow related joinpoints such as \ttt{OnTrueBranchEnter} and \ttt{OnFalseBranchEnter} which capture conditional jump branches. 
Moreover, it provides a variety of control-flow properties within the static context objects. 
For example, it is possible to traverse the CFG of a method to retrieve the successors and the predecessors of basic blocks, moreover, edges are labeled denoting if it is the True or False branch of a conditional jump.
Furthermore, BISM provides an optional feature to display the CFGs of methods before and after instrumentation, which gives developers visual assistance for analysis and insight on how to instrument the code and optimize it. 

 \paragraph*{Compatibility with ASM.}
BISM uses ASM extensively and relays all its generated class representations within the static context objects. 
% for more advanced users,
Furthermore, it allows for inserting raw bytecode instructions by using the ASM data types. 
In this case, it is the responsibility of the user to write instrumentation code free from compilation and runtime errors.  
If the user unintentionally inserts faulty instructions, the code might break. %, and the instrumentation fails. 
The ability to insert ASM instructions provides highly expressive instrumentation capabilities, especially when it comes to inlining the monitor code into the target program. 
\paragraph{Bytecode coverage.}
BISM can run in two modes: \emph{build-time} (as a standalone application) with static instrumentation, and \emph{load-time} with an agent (utilizing \ttt{java.lang} \texttt{.instrument}) that intercepts all classes loaded by the JVM and instruments before the linking phase. 
In build-time, BISM is capable of instrumenting all the compiled classes and methods\footnote{Excluding the native and abstract methods, as they do not have bytecode representation.}. 
In load-time, BISM is capable of instrumenting additional classes, including classes from the Java class library that are flagged as modifiable. 
The modifiable flag keeps certain core classes outside the scope of BISM. 
Note, modifying such classes is rather needed in dynamic program analysis (e.g., profiling, debugging). 
\section{BISM Language}
\label{sec:bism-language}
We demonstrate the language in BISM, which allows developers to write \emph{transformers} (\ie instrumentation classes). 
The language provides joinpoints (\secref{sec:joinpoints}) 
which capture exact points of program executions, 
static and dynamic contexts  (Sections~\ref{sec:staticcontext} and~\ref{sec:dynamiccontext}) 
which retrieve relevant information at joinpoints, 
and instrumentation methods (\secref{sec:instrumentationmethods}) used to instrument a target program. 
\subsection{Joinpoints}
\label{sec:joinpoints}
 
 Joinpoints identify different execution points of a program; they mark bytecode regions where instrumentation can be inlined in the target program.
BISM offers a closed set of joinpoints capable of capturing different points of a program execution, classified into three categories: Instruction, Basic Block, and Method joinpoints.
 BISM does not implement the notion of a pointcut, where a developer may select multiple joinpoints and instruments at once. 
 We list below the set of all joinpoints available and specify where the instrumented code is executed with respect to the bytecode regions. 
 \paragraph*{Instruction joinpoints.} BISM provides the following instruction-related joinpoints:   
    \begin{itemize}
        \item   \ttt{BeforeInstruction}: captures execution before a bytecode instruction. If the instruction is the entry point of a basic block, the code executes after the instruction. 
        \item
        \ttt{AfterInstruction}: captures execution after a bytecode instruction. If the instruction is the exit point of a basic block, the code executes before thee instruction.  
        \item
        \ttt{BeforeMethodCall}: captures execution before a method call instruction and after loading all needed values on the stack. 
        \item 
        \ttt{AfterMethodCall}: captures execution immediately after a method call instruction and before storing the return value from the stack, if any. 
    \end{itemize}
\paragraph*{Method joinpoints.} BSIM also provides two method-related joinpoints: 
    \begin{itemize}
        \item  \ttt{OnMethodEnter}: captures execution on method entry block, same execution rules as OnBasicBlockEnter. 
        \item 
        \ttt{OnMethodExit}: captures execution on all exit blocks of a method before the return or throw instruction.
    \end{itemize}
\paragraph*{Basic block joinpoints.} In addition to the previous joinpoints, BISM provides basic block-related joinpoints which facilities instrumenting for control-flow analysis:  
    \begin{itemize}
        \item  \ttt{OnBasicBlockEnter}: captures execution at the entry of a basic block, at the first real 
        instruction\footnote{Real instructions are instructions that actually get executed, as opposite to some special Java bytecode instructions such as Label and Line number instructions.}.
        \item 
        \ttt{OnBasicBlockExit}: captures execution after the last instruction of a basic block; except when last instruction is a JUMP/RETURN/THROW instruction, then it executes before the last instruction. 
        \item 
        \ttt{OnTrueBranchEnter}: captures execution on the entry of a successor block after a conditional jump on True evaluation. 
        \item 
        \ttt{OnFalseBranchEnter}: captures execution on the entry of a successor block after a conditional jump on False evaluation. 
    \end{itemize}
 The order of which joinpoints are visited first when entering a method is as follows: OnMethodEnter, 
 OnBasicBlockEnter, OnTrueBranchEnter, OnFalseBranchEnter, BeforeInstruction, 
 BeforeMethodCall, AfterMethodCall, AfterInstruction, OnBasicBlockExit, OnMethodExit. 

 \subsection{Static Context}
 \label{sec:staticcontext}
Static context objects provide relevant static information at joinpoints.  
These objects can be used to retrieve information about a bytecode instruction, a method call, a basic block, a method, and a class.
BISM performs static analysis on target programs and provides additional control-flow-related static information such as basic block successors and predecessors. 
We list all the static context objects available and their properties. 

\paragraph*{Instruction context.} The \ttt{Instruction} context 
 provides all relevant information about a single instruction being visited, and it contains the following fields: 
 \begin{itemize}
    \item \ttt{index}: a unique instruction index. 
     \item \ttt{node}: the ASM org.objectweb.asm.tree.AbstractInsNode that can be casted into a more specific AbstractInsNode sub type. 
     \item \ttt{opcode}: the bytecode instruction opcode. 
     \item \ttt{next}: the next instruction in the current basic block. Null if at the end of a basic block. 
     \item \ttt{previous}: the previous instruction in the basic block. Null if at the beginning of a basic block. 
     \item \ttt{isConditionalJump()}: true if instruction is a conditional jump instruction. 
     \item \ttt{isBranchingInstruction()}: true if instruction is an instance of JumpInsnNode, LookupSwitchInsnNode, TableSwitchInsnNode, or opcode is ATHROW, RET, IRETURN or RETURN. 
     \item \ttt{stackOperandsCountIfConditionalJump()}: the number of stack opera\-nds a conditional jump consumes. Equal to -1 in the case of non-conditional jumps. 
     \item \ttt{getBasicValueFrame()}: contains a list of all local variables, stack items, and their types at the stack frame before executing the current instruction. 
     \item \ttt{getSourceValueFrame()}: contains a list of all local variables and stack items and their source i.e. what instruction created/manipulated them. 
     \item \ttt{methodName}: the method name; the owner of the current instruction. 
     \item \ttt{basicBlock}: the \ttt{BasicBlock} context of the current instruction. 
     \item \ttt{className}: the name of the class; the owner of the current  instruction. 
 \end{itemize}

\paragraph*{MethodCall context.} A special type of \ttt{Instruction} context (only available before and after method calls). 
In addition to its \ttt{Instruction} context, it provides the following fields: 
 \begin{itemize}
     \item \ttt{methodOwner}: the name of the called class (callee). 
     \item \ttt{methodName}: the name of the method called. 
     \item \ttt{currentClassName}: the name of the calling class. 
     \item \ttt{node}: the ASM MethodInsnNode instruction. 
     \item \ttt{ins}: references the instruction. 
 \end{itemize}

 \paragraph*{BasicBlock context.} The \ttt{BasicBlock} context provides information about the current basic block being visited, 
 and it contains the following fields:
 \begin{itemize}
     \item \ttt{id}: a unique String that identifies the basic block. 
     \item \ttt{index}: a unique index that identifies the basic block inside  a class.
     \item \ttt{blockType}: the block type, which can be Normal, ConditionalJump, Goto, Switch, or Return. 
     \item \ttt{size}: the number of instructions in the basic block. 
     \item \ttt{getSuccessorBlocks()}: all successors of the basic block as per the CFG. 
     \item \ttt{getPredecessorBlocks()}: all predecessors of the basic block as per the CFG. 
     \item \ttt{getTrueBranch()}: the target block after a conditional jump evaluates to true, is null if the block does not end with a conditional jump. 
     \item \ttt{getFalseBranch()}: the target block after a conditional jump evaluates to false, is null if the block does not end with a conditional jump. 
      \item \ttt{getFirstInstruction()}: the first AbstractInsNode in the  basic block. 
     \item \ttt{getFirstRealInstruction()}: the first real instruction in the  basic block. 
     \item \ttt{getLastRealInstruction()}: the last real instruction in the  basic block. 
     \item \ttt{method}: the \ttt{Method} context of the basic block. 
 \end{itemize}

 \paragraph*{Method context.} The \ttt{Method} context provides info about the method being visited and has the following fields:
 \begin{itemize}
     \item \ttt{name}: the name of the method. 
     \item \ttt{methodNode}: the ASM org.objectweb.asm.tree.MethodNode. 
     \item \ttt{getNumberOfBasicBlocks()}: the number of basic blocks in the method. 
     \item \ttt{getEntryBlock()}: the entry basic block. 
     \item \ttt{getExitBlocks()}: a list of all exiting basic blocks. 
     \item \ttt{classContext}: the \ttt{Class} context of the method. 
 \end{itemize}

 \paragraph*{Class context.} The \ttt{Class} context provides information about the class being instrumented and has the following fields:
 \begin{itemize}
     \item \ttt{name}: the name of the method. 
     \item \ttt{classNode}: the ASM org.objectweb.asm.tree.ClassNode
 \end{itemize}
Static contexts are composed in a hierarchical structure such that an \ttt{Instruction} context object contains a reference to its \ttt{BasicBlock}  context, \ttt{BasicBlock} context to its \ttt{Method} context, and a \ttt{Method} context to its \ttt{Class} context. 
 
\begin{lstlisting}[style=JavaStyle,
      backgroundcolor=\color{white}, 
      caption={A transformer for intercepting basic block executions.}, 
    captionpos=b,
    label=lst:staticcontextexample,
    numbers=none,
    float=hbtp,
    frame=bt
  ]
public class BasicBlockTransformer extends Transformer {

    @Override
    public void onBasicBlockEnter(BasicBlock bb){
        String blockId = bb.method.className+"."+bb.method.name+"."+bb.id;

        print("Entered block:" + blockId)
    }

    @Override
    public void onBasicBlockExit(BasicBlock bb)
        String blockId = bb.method.className+"."+bb.method.name+"."+bb.id;

        print("Exited block:" + blockId)
    }
}
\end{lstlisting}
The transformer depicted in Listing~\ref{lst:staticcontextexample} uses the  joinpoints \ttt{onBasicBlockEnter} and \ttt{onBasicBlockExit} to intercept all basic block executions. % in a program execution.  
The static context \ttt{BasicBlock bb} is used to get the block id, the method name, and the class name. 
Here, the instrumentation method \ttt{print} inserts a print invocation in the target program before and after every basic block execution.

% \subsection{Scope}
% \label{sec:scopes}

% BISM provides features to limit the scope of instrumentation.
% %
% The \emph{scope} argument can be specified when running BISM to match classes and methods by their names.
% %
% Specifying \emph{(scope=java.util.List.*, java.util.Iterator.next)} will instrument all methods in the \ttt{List} class and only the next method in the \ttt{Iterator} class.
% %
% Also, specifying package names limits the scope for packages.
% %
% Inside joinpoints, static context objects can also be used to limit the scope of instrumentation.
% %
% Listing \ref{lst:dynamiccontextexample}, shows an example of using the static context object \ttt{MethodCall mc} to filter for a specific method by its name and class owner.

\subsection{Dynamic Context}
\label{sec:dynamiccontext}
In addition to static context, BISM provides dynamic context objects at all joinpoints.
These objects are capable of accessing dynamic values that are possibly only known during the target program execution.
Dynamic Context objects provide access to dynamic values that are possibly only known during execution.
BISM gathers this information from local variables and operand stack, then weaves the necessary code to extract this information. 
In some cases (e.g., when accessing stack values), BISM might instrument additional local variables to store them for later use.
We list the methods available in dynamic contexts; note all these calls return a \ttt{DynamicValue} object omitted for brevity: 

\begin{itemize}
    \item \ttt{getThis()}: returns a reference to the class owner of the method being instrumented, and null if the class or method is static.  
    \item \ttt{getLocalVariable(int)}: returns a reference to a local variable by index. 
    \item \ttt{getStackValue(int)}: returns a reference to a stack value. 
    \item \ttt{getInstanceField(String)}: returns a reference to an instance field in the class being instrumented, and null if static. 
    \item \ttt{getInstanceField(DynamicValue, String, Class)}: returns a reference to an instance field in a \ttt{DynamicValue}, and null if field is static or the dynamic value is not an object.
    \item \ttt{getStaticField(String)}: returns a reference to a static field in the class being instrumented. 
    \item \ttt{getStaticField(DynamicValue, String, Class)}: returns a reference to a static field in a \ttt{DynamicValue}, and null if the dynamic value is not an object. 
\end{itemize}
Additionally, we list the values related to these methods: %the below methods are only available at joinpoints related to methods and method calls.
\begin{itemize}
    \item \ttt{getMethodArgs(int)}: returns a reference to a method argument by index starting at 1. Only available in \ttt{MethodCall} and \ttt{Method} joinpoints.
    \item  \ttt{getMethodReceiver()}: returns a reference to the object whose method is being called. Returns null for static methods. Only available only in \ttt{MethodCall} joinpoints.
    \item \ttt{getMethodResult()}: returns a reference to a method result. Only available only in \ttt{MethodCall} joinpoints.
\end{itemize}
BISM also allows to add new local variables to a method explicitly; these are useful for different purposes like to pass data across joinpoints.
Note that the scope of the values of these variables is the method where they are created. 
\begin{itemize}
    \item \ttt{addLocalVariable(Object value)}: creates a new local variable and sets it to a primitive value, then return its reference as a 
    \ttt{LocalVariable} type. 
    This is only available in \ttt{Method} joinpoints. 
    \item \ttt{updateLocalVariable(LocalVariable, Object value)}: updates a \ttt{LocalVariable} and sets it to a primitive value. This is available in all joinpoints.
\end{itemize}
Listing~\ref{lst:dynamiccontextexample} presents a transformer using \ttt{afterMethodCall} joinpoint to capture the return of an \ttt{Iterator} created from a \ttt{List} object,
 and retrieving dynamic data from the dynamic context object \ttt{MethodCallDynamicContext dc}.  The example also shows how to limit the scope using an if-statement to a specific method.
 Note that BISM also provides a more general \emph{scope} argument that can be specified at runtime to match packages, classes, and methods by names (using possibly wildcards). 

\begin{lstlisting}[style=JavaStyle,
    %linewidth=0.8\textwidth,
    backgroundcolor=\color{white},   
    caption={A transformer that intercepts the creation of an iterator from a \ttt{List}.
    },
    captionpos=b,
    label=lst:dynamiccontextexample,
    numbers=none,
    float=htb,
    frame=bt
  ]
public class IteratorTransformer extends Transformer {

 @Override
 public void afterMethodCall(MethodCall mc, MethodCallDynamicContext dc){

    if (mc.methodName.equals("iterator") && mc.methodOwner.endsWith("List")) {

        // Access to dynamic data
        DynamicValue callingClass = dc.getThis(mc);
        DynamicValue list = dc.getMethodTarget(mc);
        DynamicValue iterator = dc.getMethodResult(mc);
        
        // Invoking a monitor
        StaticInvocation sti = 
            new StaticInvocation("IteratorMonitor", "iteratorCreation");
        sti.addParameter(callingClass);
        sti.addParameter(list);
        sti.addParameter(iterator);
        invoke(sti);
    }
 }
}
\end{lstlisting}

\subsection{Instrumentation methods}
\label{sec:instrumentationmethods}

% Instrumentation logic is specified in BISM by a developer in a single Java class of type \ttt{Transformer}. 
% %
% In this class, the developer selects the joinpoints, and as seen previously, static and dynamic context objects can be retrieved and manipulated.
% %
% To instrument the code, a developer needs to invoke special instrumentation methods.
% %
% These methods are eventually compiled by BISM into bytecode instructions and inlined at the appropriate locations.
% %
% We list below the instrumentation methods available in BISM:

A developer instruments the target program using specified instrumentation methods.
BISM provides \ttt{print} methods with multiple options to invoke a print command.
It also provides (i) \ttt{invoke} methods for static method invocation and (ii) \ttt{insert} methods for bytecode instruction insertion.
These methods are compiled by BISM into bytecode instructions and inlined at the exact joinpoint locations. 
%
% For example, Listing~\ref{lst:staticcontextexample} shows the use of \ttt{print} to print the constructed id of a basic block. 
%
% Listing~\ref{lst:dynamiccontextexample} shows how a method invocation is instrumented after a method call.
%
We list below the instrumentation methods available in BISM. 

\paragraph*{Printing on console.}
Instrumenting print statements in the target program can be achieved via multiple print instrumentation methods available in BISM.
These methods take either static values or dynamic values retrieved in joinpoints.
Listing \ref{lst:staticcontextexample} shows an example of using one of the print helper methods to instrument the target program to print the basic block constructed id. 
We list all of these methods: 
\begin{itemize}
    \item 
    \ttt{print(String)}: prints a message on the console. 
    \item 
    \ttt{println(String)}: prints a message on the console followed by a new line.  
    \item 
    \ttt{print(DynamicValue)}: prints the \ttt{toString()} of a dynamic value on the console. 
    \item 
    \ttt{printHash(DynamicValue)}:  
    prints the unique identity hash code of a dynamic value. % \ttt{System.identityHashCode()} of a dynamic value. 
    \item 
    \ttt{print(String, boolean)}: similar to \ttt{print(String)} but if passed bool\-ean true, the print stream will be \ttt{err}. 
    \item 
    \ttt{print(DynamicValue, boolean)}: similar to \ttt{print(DynamicValue)} but if passed boolean true, the print stream will be \ttt{err}. 
    \item 
    \ttt{printHash(DynamicValue, boolean)}:  
    similar to \ttt{printHash(Dynami}\-\ttt{cValue)} but if passed boolean true, the print stream will be \ttt{err}.
\end{itemize}

\paragraph*{Invoking static methods.}  
Invoking external static methods can be achieved using the instrumentation method \ttt{invoke}. 
An object of type \ttt{StaticInvocation} should be constructed in a joinpoint and provided with the external class name,the method name, and parameters.
%
% The method signature is: 
% %
% \ttt{invoke(StaticInvocation smi)}  
%
Listing \ref{lst:dynamiccontextexample} depicts a transformer that instrument  the target program to call an external monitor method \ttt{iteratorCreation}.
The \ttt{StaticInvocation} constructor takes in the monitor class name and method name.
\ttt{addParameter()} is then called to add parameters to the invocation, 
it supports either \ttt{DynamicValue} type, or any primitive type in Java including \ttt{String} type (any other type will be ignored). 
After that, \ttt{invoke} weaves the method call in the target program.

\paragraph*{Raw bytecode instructions.} 
Inserting raw bytecode instructions can be achieved using two insert methods, 
one takes as an argument a single ASM \ttt{AbstractInsnNode} instruction, and the other takes a list of instructions.
When using these methods, it is the responsibility of the developer to write correct instructions and avoid breaking the code. 
Errors can be introduced by ignoring the stack requirements and altering local variables.
For Java 8 and above programs, using the insert methods to push new values on the stack or create local variables requires modifying
the maxStack and maxLocals values.
All static context objects, hold the needed ASM object \ttt{MethodNode} to increment the values maxLocals and maxStack from within the joinpoint.
These methods are: 
\begin{itemize}
\item \ttt{insert(AbstractInsnNode ins)}  
\item \ttt{insert(List<AbstractInsnNode> ins)}
\end{itemize}

%
%==========================================================
\section{BISM Implementation}
\label{sec:framework}
%==========================================================

\begin{figure}
\centering
	\includegraphics[width=0.72\textwidth]{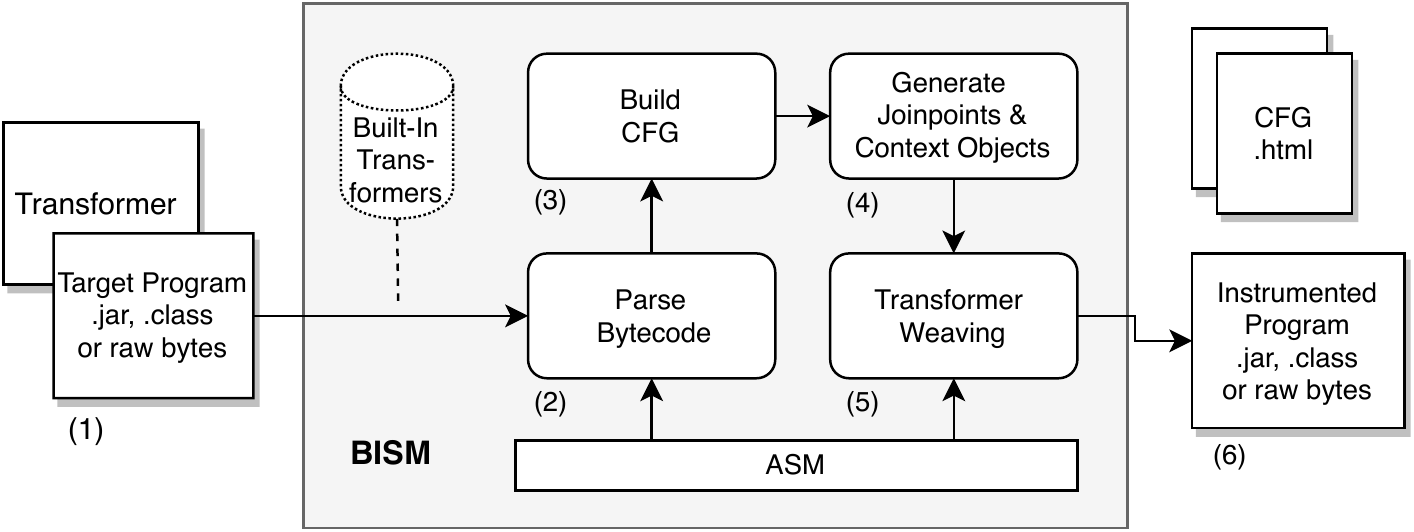} 
	\caption{Instrumentation process in BISM.} 
	\label{fig:framework-overview}
\end{figure}

% In this section, we provide details about BISM implementation. 
% %
% In Section~\ref{sec:framework-architecture}, we describe the general architecture of BISM.
% %
% In Section~\ref{sec:framework-steps}, we demonstrate the steps involved in the instrumentation process. 
% %
% %----------------------------------------------------------
% \subsection{Architecture}
% \label{sec:framework-architecture}
% %----------------------------------------------------------
BISM is an open-source tool \cite{bism} implemented in Java using about 4,000 LOC and 40 classes distributed in separate modules. 
It uses ASM for bytecode parsing, analysis, and weaving. 
%
%BISM can run in two modes: a build-time mode where BISM runs as a standalone application to statically instrument a program, and a load-time mode where BISM is attached to a program as a Java agent . 
%
Fig.~\ref{fig:framework-overview} shows BISM internal workflow.\\
%, called an \emph{instrumentation run}. 
%
% \begin{figure}[htbp]
%   \centering
%   \includegraphics[width=0.9\textwidth]{figures/bism-architecturenew.pdf} 
%   \caption{Instrumentation process in BISM.} 
%   \label{fig:framework-overview}
% \end{figure}
%
% \begin{itemize}
%
% \begin{wrapfigure}[10]{r}{8.6cm}
% \end{wrapfigure}
%--------------------------------
% \item 
\textbf{(1) User Input.}
%--------------------------------
In build-time mode, BISM takes a target program bytecode (\emph{.class} or \emph{.jar}) to be instrumented, and a transformer which specifies the instrumentation logic. 
In load-time mode, BISM only takes a transformer, which is used to instrument every class being loaded by the JVM. 
BISM provides several built-in transformers that can be directly used.   %specified by their name.
Moreover, users can specify a scope to filter target packages, classes, or methods. \\%, and can enable storing CFGs into html files. 
%
%--------------------------------
% \item 
\textbf{(2) Parse Bytecode.}
%--------------------------------
BISM uses ASM to parse the bytecode and to generate a tree object which contains all the class details, such as fields, methods, and instructions. \\
%
%----------------------------
% \item 
\textbf{(3) Build CFG.}
%----------------------------
BISM constructs the CFGs for all methods in the target class.  
If the transformer utilizes control-flow joinpoints (\ttt{onTrueBranch} and \ttt{onFalseBranch}), BISM eliminates all \emph{critical edges} from the CFGs to avoid instrumentation errors. 
This is done by inserting empty basic blocks in the middle of critical edges. 
Note, BISM keeps copies of the original CFGs. 
Users can optionally enable the \emph{visualizer} to store CFGs in HTML files on the disk. \\
%-------------------------------
% \item 
\textbf{(4) Generate Joinpoints and Context Objects.}
%-------------------------------
%
BISM iterates over the target class to generate all joinpoints utilizing the created CFGs.
At each joinpoint, the relevant static and dynamic context objects are created. \\ 
%
%-------------------------------
% \item 
\textbf{(5) Transformer Weaving.}
%-------------------------------
BISM evaluates the used dynamic contexts based on the joinpoint static information and weaves the bytecode needed to extract concrete values from executions. %the target program. 
It then weaves instrumentation methods by compiling them into bytecode instructions that are woven into the target program at the specified joinpoint.  \\
%
%-------------------------------
% \item 
\textbf{(6) Output.}
%-------------------------------
The instrumented bytecode is then output back as a \emph{.class} file in build-time mode, or passed as raw bytes to the JVM in load-time mode. 
In case of instrumentation errors, e.g., due to adding manual ASM instructions, BISM  emits a weaving error.  
If the visualizer is enabled, instrumented CFGs are stored in HTML files on the disk. 
%
% \end{itemize}
%
\section{Evaluation}
\label{sec:casestudies}
We compare BISM with DiSL using two complementary experiments. 
To guarantee fairness, we switched off adding exception handlers around instrumented code in DiSL. 
% In the first two experiments AES (Advanced Encryption Standard) and a financial transaction system are statically instrumented using build-time mode to, respectively, detect test inversions and arbitrary jumps. % in the control-flow of executions.  
% %
% The third example demonstrates a general runtime verification case where seven applications from the DaCapo benchmark~\cite{DaCapo06} are instrumented using both build-time and load-time modes to verify the classical \textbf{HasNext}, \textbf{UnsafeIterator} and \textbf{SafeSyncMap} properties\footnote{
% 	\textbf{HasNext} property specifies that a program should always call \emph{hasNext()} before calling \emph{next()} on an iterator.
% 	%
% 	\textbf{UnSafeIterator} property specifies that a collection should not be updated when an iterator associated with it is being used.
% 	%
% 	\textbf{SafeSyncMap} property specifies that a map should not be updated when an iterator associated with it is being used.
% 	}. \todo{Is the footnote really necessary?}
% %
%
In what follows, we illustrate how we carried out our experiments. % and the obtained results. 
% %

%
%
\subsection{Inline Monitor to Detect Test Inversions}
\label{sec:aes}
We instrument an external AES (Advanced Encryption Standard) implementation in build-time mode to detect test inversions.  % in the control-flow of executions.  
The instrumentation deploys inline monitors that duplicate all conditional jumps 
in their successor blocks to report test inversions.
%
% The monitor also reports an error in case of any inconsistency. 
%
%In BISM, we use built-in features to duplicate conditional jumps utilizing \ttt{insert} instrumentation method to add raw bytecode instructions.
%
% The monitor also reports an error in case of any inconsistency. 
%
In \textbf{BSIM}, we use the \ttt{beforeInstruction} joinpoint to capture all conditional jumps.
We extract the opcode from the static context object \ttt{Instructions} and use the instrumentation method \ttt{insert} to duplicate the jump-related stack values
\footnote{Note that extracting stack values can be also achieved using dynamic context method \ttt{getStackValue} and adding new local variables.}.
We then use the control-flow joinpoints \ttt{OnTrueBranchEnter} and \ttt{onFalseBranchEnter} to capture the blocks executing after the jump.
We inline at the beginning of these blocks, utilizing \ttt{insert}, a duplicate test that reports any inconsistency.
This test is written in bytecode instructions based on the last captured conditional jump, and reports a test inversion attack if it happens.
%
%
% \begin{wraptable}{r}{3.5cm}
% \caption{Bytecode size.}
% \label{tab:aes-bytecodesize}
%   \centering
%   \begin{tabular}{c|c} 
%   %
%   \hline
%   \rowcolor{gray!30}
%                      & Size (KB) \\ 
%   \hline
%   Original & 9 \\
%   BISM     & 10 ($\times$1.11) \\
%   DiSL     & 128 ($\times$14.22) \\
%   \hline
%   \end{tabular}
% \end{wraptable} 
% the static context \ttt{Instruction} is used at the joinpoint \ttt{beforeInstruction} to spot conditional jumps and retrieve the number of operands it consumes. 
% %
% Then, at the joinpoint \ttt{beforeInstruction}, the instrumentation method \ttt{insert} is used to insert instruction \ttt{DUP} or \ttt{DUP2} to duplicate the relevant stake values (depending on the number of operands). 
% %
% Finally, at both joinpoints \ttt{OnTrueBranchEnter} and \ttt{onFalseBranchEnter}, a new \ttt{LabelNode} is created, and \ttt{insert} is used to reproduce the conditional jump instruction. 
% %
% In the case of inconsistency in the checks' results, the instrumentation method \ttt{print} is used to report an error. Otherwise, the execution continues normally. 
%
% And, on successor blocks, we map opcodes to Java syntax to re-evaluate conditional jumps using switch statements.
%
In \textbf{DiSL}, we write multiple instrumentation snippets, using the \ttt{BytecodeMarker} to capture all conditional jumps before they occur. 
We implement a custom \ttt{InstructionStaticContext} object to retrieve additional static information from conditional jump instructions 
such as the index of a jump target. 
We also use the dynamic context object to retrieve stack values.
We then store the extracted information in synthetic local variables, and we add a flag to specify that a jump has occurred.
Using the \ttt{BasicBlockMarker}, we capture basic block entries and check if a jump occurred before entering each block.
Accordingly, we re-evaluate the conditional jump in Java syntax using a switch statement on opcodes and the expected target.
Hence, we report any inconsistency if it happened.
%
% \begin{figure}[h]
%     \subfloat[Runtime.]{
%       \begin{minipage}[c][1\width]{
%          0.5\textwidth}  
%          \centering  
%          \includegraphics[width=1\textwidth]{figures/aestime.pdf}
%       \end{minipage}}  
%    \hfill   
%     \subfloat[Memory footprint.]{
%       \begin{minipage}[c][1\width]{     
%          0.5\textwidth}    
%          \centering
%          \includegraphics[width=1\textwidth]{figures/aesmemory.pdf}
%       \end{minipage}}
%   \caption{Runtime and memory footprint by AES on files of different sizes.}
%   \label{fig:aes}
%   \end{figure} 

\begin{figure}[t]
\vspace*{-12pt}
\centering
  \begin{subfigure}[b]{0.495\textwidth}
    \includegraphics[width=\textwidth]{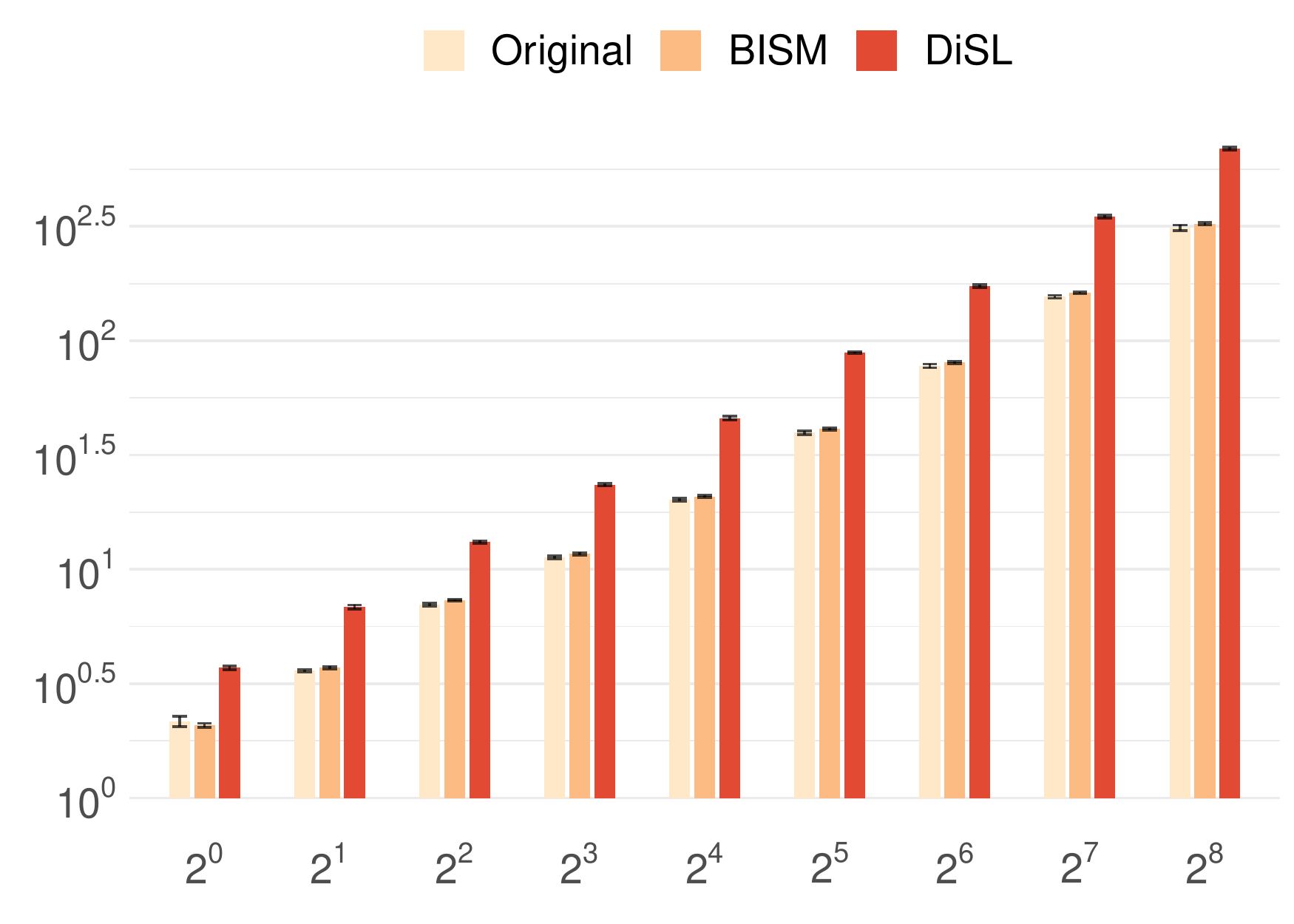}
    \caption{Runtime (ms).}
  \end{subfigure}
  \hfill
  \begin{subfigure}[b]{0.495\textwidth}
    \includegraphics[width=\textwidth]{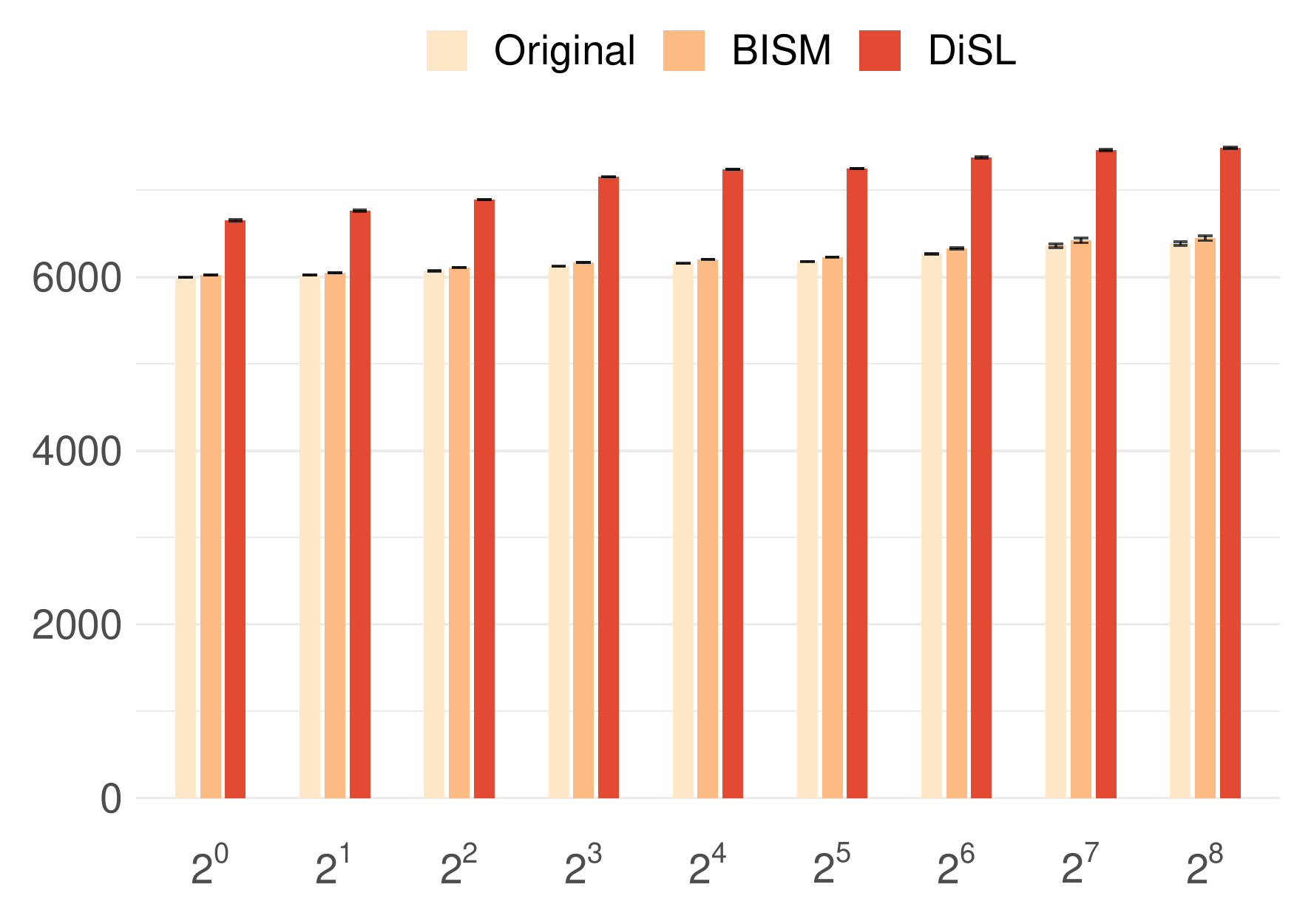}
    \caption{Memory footprint (KB).}
  \end{subfigure}
  \caption{Runtime and memory footprint by AES on files of different sizes.}
  \label{fig:aes}
\end{figure}
We use AES to encrypt then decrypt input files of different sizes, line by line.  
The bytecode size of the original AES class is 9 KB. 
%
% After instrumentation using BISM and DiSL it is 10 KB ($\times$1.11) and 128 KB ($\times$14.22), respectively. 
%
After instrumentation, it is 10 KB (+11.11\%) for BISM, and 128 KB (+1322\%) for DiSL. 
The significant overhead in DiSL is due to the inability to inline the monitor in bytecode and having to instrument it in Java.
% %
% We also had to inline the monitors in all basic blocks, even though some of these monitors never executes.  
% %
% In contrast to BISM (\ttt{OnTrueBranchEnter} and \ttt{onFalseBranchEnter}), DiSL does not provide out-of-the-box control-flow information in custom markers API to target the branching blocks only.
% %
Fig.~\ref{fig:aes} reports runtime and memory footprint with respect to file size (KB). 
For each input file, we performed 100 measurements and reported the mean and the standard deviation.
The latter is very low.  
We used Java JDK 8u181 with 4 GB maximum heap size on a standard PC (Intel Core i7 2.2 GHz, 16 GB RAM) running macOS Catalina v10.15.5 64-bit. 
The results show that BISM incurs less overhead than DiSL for all file sizes.  
Table~\ref{tab:aes} reports the number of events (\ie checks duplicated). %conditional jump executions duplicated). 
%
% The experiment was conducted using Java JDK 8u181 with 4 GB maximum heap size on a standard PC (Intel Core i7 2.2 GHz, 16 GB RAM) running macOS Catalina v10.15.5 64-bit. 
%
\bgroup
\setlength{\tabcolsep}{5pt}
\begin{table}[]
\caption{Number of events by AES class (in millions).}
\label{tab:aes}
  \centering
  \begin{tabular}{c|c|c|c|c|c|c|c|c|c} 
  %
  % \hline 
  % \rowcolor{gray!30}
  Input File (KB) & $2^0$ & $2^1$ & $2^2$ & $2^3$ & $2^4$ & $2^5$ & $2^6$ & $2^7$ & $2^8$\\ 
  \hline

  Events (M) & 0.92 & 1.82 & 3.65 & 7.34 & 14.94 & 29.53 & 58.50 & 117.24 &  233.10\\ 
  % 926861 1827047 3652915 7343485 14944523 29532229 58509341 117245429 233109037
  % \hline
  % \hline 
  % \rowcolor{gray!30}
  %                    & \multicolumn{3}{c|}{Original} & \multicolumn{3}{c|}{BISM} & \multicolumn{3}{c}{DiSL} \\
  % \hline
  % Bytecode Size (KB) & \multicolumn{3}{c|}{9} & \multicolumn{3}{c|}{10 ($\times$1.11)} & \multicolumn{3}{c}{128 ($\times$14.22)} \\
  % \hline
  \end{tabular}
  \vspace*{-.5cm}
\end{table}
\egroup
\subsection{DaCapo Benchmarks}
\label{sec:dacapo}

We compare BISM, DiSL and AspectJ in a general runtime verification scenario. 
We use HasNext, UnSafeIterator and SafeSyncMap properties. 
HasNext property specifies that a program should always call \emph{hasNext()} before calling \emph{next()} on an iterator.
UnSafeIterator property specifies that a collection should not be updated when an iterator associated with it is being used. 
SafeSyncMap property specifies that a map should not be updated when an iterator associated with it is being used.
%
% We limit our scope to method calls to \ttt{java.util} types \ttt{List}, \ttt{Collection}, \ttt{Iterator} and \ttt{Map} only.
%
We instrument, in build-time and load-time mode, the benchmarks in the DaCapo suite~\cite{DaCapo06} (dacapo-9.12-bach), targeting only the packages specific to each benchmark. 
We implement an external monitor library to receive the events with methods that only count the number of invocations.
Instrumentation in BISM is straightforward and written in one Transformer class.
%
% To instrument in BISM, we use the static context provided at method call joinpoints to filter methods by their names and owners. 
% %
% To access the receivers and results of the method calls, we utilize the methods available in dynamic contexts. 
% %
Method calls are captured using joinpoints \ttt{beforeMethodCall} and \ttt{afterMethodCall}, and are filtered by their names and owners 
using the static context object provided with the joinpoints.
To limit the scope to the specific benchmark packages, we use the runtime argument \emph{scope}.
To access the receivers and results of the method calls (dynamic values), we 
utilize \ttt{getMethodReceiver()} and \ttt{getMethodResult()} methods.
After that, we construct a \ttt{StaticInvocation} object and add the dynamic values to this object.
Then to invoke the external monitor library, we utilize \ttt{invoke} instrumentation methods.
% In DiSL, we implement custom Markers to capture the needed method calls and use argument processors and dynamic context objects to access dynamic values. 
 Instrumentation in DiSL is written in several classes. 
For each different method call, we write a custom \ttt{Marker} that filters using ASM syntax for a method name and owner.
For each marker, we implement an instrumentation snippet in the main instrumentation class.
To limit the scope to certain packages, we use the \ttt{scope} annotation on each instrumentation snippet.
To access the receivers of the method calls, we use argument processors, and to access method results we use
their dynamic context method \ttt{getStackValue()}.
Then to invoke the external monitors, we include the monitor library in the instrumentation package and call the methods directly.
Instrumentation in AspectJ is written in one Aspect class.
We capture method calls with \ttt{call} joinpoints.
We defined respective pointcuts and advices to invoke the monitors. 
We limit the scope to specific packages in the configuration file.

%
% In build-time, we write instrumented classes on the disk and merge them with the original files in Dacapo to have statically instrumented programs.
% %
% The measurement machine is an Intel i9-9980HK Core2 (2.4 GHz, 8 GB RAM) running Ubuntu 20.04 LTS 64-bit.
%
% We use Oracle’s JDK 1.8.0\_251 Hotspot 64-Bit Server VM with 2 GB maximum heap size.
%
 
% \begin{figure}[h]
%     \subfloat[Build-time]{
%       \begin{minipage}[c][1\width]{
%          0.5\textwidth}  
%          \centering  
%          \includegraphics[width=1\textwidth]{figures/dacapobuildtime.pdf}
%       \end{minipage}}  
%    \hfill   
%     \subfloat[Load-time]{
%       \begin{minipage}[c][1\width]{     
%          0.5\textwidth}    
%          \centering
%          \includegraphics[width=1\textwidth]{figures/dacapoloadtime.pdf}
%       \end{minipage}}
%   \caption{Dacapo benchmarks execution time in ms}
%   \label{fig:dacapoexecutiontime}
%   \end{figure} 

\begin{figure}[htbp]
\vspace*{-15pt}
  \begin{subfigure}[b]{0.5\textwidth}
    \includegraphics[width=\textwidth]{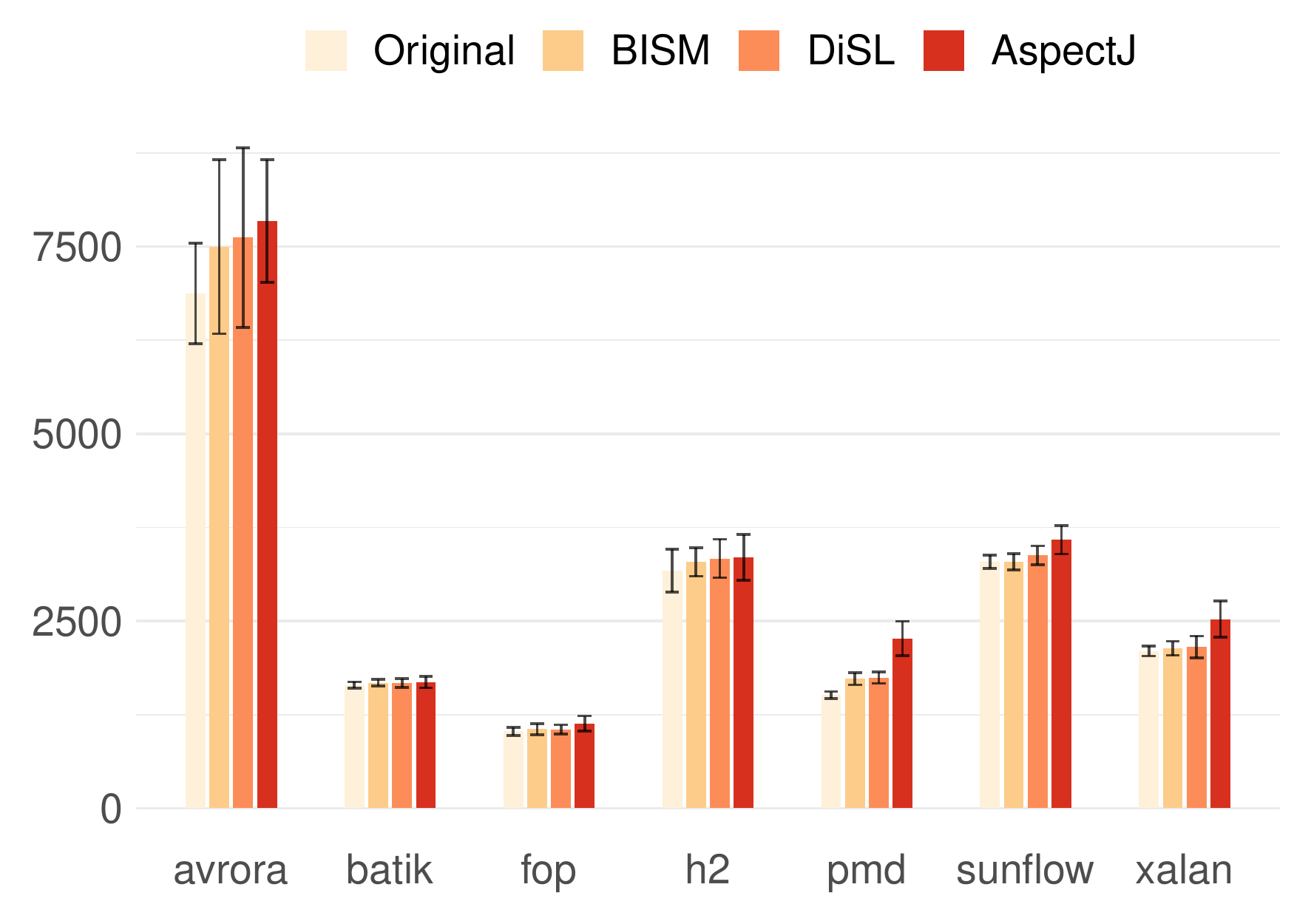}
    \caption{Build-time.}
  \end{subfigure}
  \hfill
  \begin{subfigure}[b]{0.5\textwidth}
    \includegraphics[width=\textwidth]{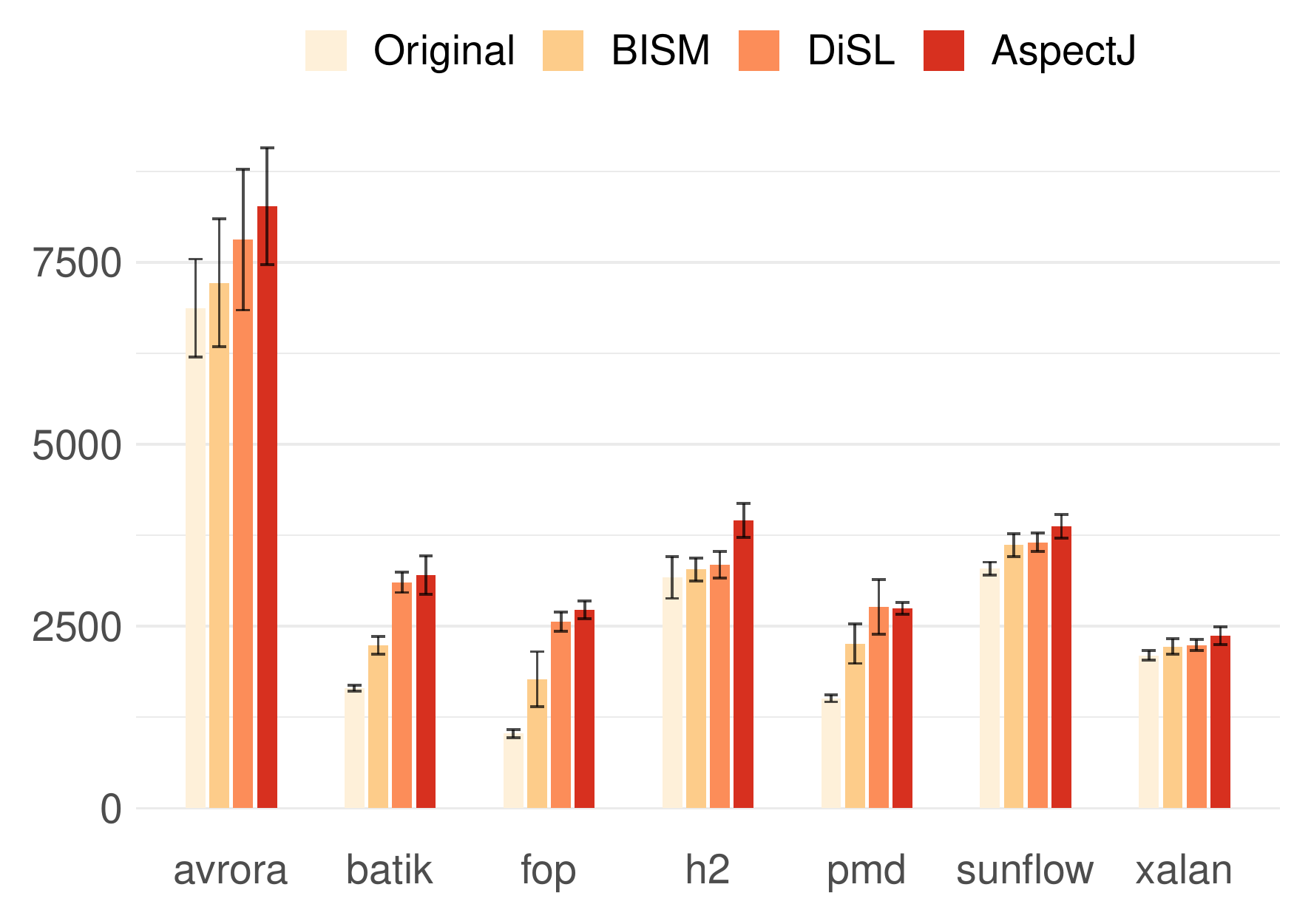}
    \caption{Load-time.}
  \end{subfigure}
  \caption{DaCapo benchmarks execution time in ms.}
  \label{fig:dacapoexecutiontime}
\end{figure}

We used Java JDK 8u251 with 2 GB maximum heap size on an Intel Core i9-9980HK (2.4 GHz, 8 GB RAM) running Ubuntu 20.04 LTS 64-bit.
Fig.~\ref{fig:dacapoexecutiontime} reports the runtime for running BISM and DiSL in (a) build-time and (b) load-time.
Our measurements correspond to the mean of 15 runs on each benchmark, also showing the standard deviation. 
For build-time mode, BISM instrumentation shows less overhead on average.
For load-time mode, BISM shows less overhead in all benchmarks.
Table \ref{tab:dacapo-size} %provides more details in terms of bytecode size.
%
% We 
reports the number of emitted events in each benchmark\footnote{The number of emitted events matches between BISM and DiSL. Even with non-determinism in specific benchmarks, the variation in the number of events is negligible.} % s are in the same range.}, 
 the number of classes in the scope of instrumentation, instrumented classes (Ins.), and their bytecode size with overhead percentages (Ovh.).
The results show that BISM incurs less overhead in all benchmarks.
\bgroup
\setlength{\tabcolsep}{4pt}
\begin{table}[htbp]
\caption{Bytecode size of the instrumented benchmarks applications.}
    \label{tab:dacapo-size}
   \centering
 
   \begin{tabular}{rllccccccccccccccccccccl}
    &                       &  & \multicolumn{1}{r}{}       & \multicolumn{1}{l}{}  & \multicolumn{1}{l}{} & \multicolumn{1}{r}{}      & \multicolumn{1}{l}{}  & \multicolumn{1}{l}{} & \multicolumn{1}{r}{}     & \multicolumn{1}{l}{}  & \multicolumn{1}{l}{} &          & \multicolumn{1}{l}{}  & \multicolumn{1}{l}{} & \multicolumn{1}{l}{} & \multicolumn{1}{l}{} & \multicolumn{1}{l}{} & \multicolumn{1}{l}{}  & \multicolumn{1}{l}{} & \multicolumn{1}{l}{} & \multicolumn{1}{l}{} & \multicolumn{1}{l}{} &  \\
    & \multicolumn{1}{l|}{} &  & \multicolumn{1}{r}{Events} & \multicolumn{1}{l|}{} & \multicolumn{1}{l}{} & \multicolumn{1}{r}{Scope} & \multicolumn{1}{l|}{} & \multicolumn{1}{l}{} & \multicolumn{1}{r}{Ins.} & \multicolumn{1}{l|}{} & \multicolumn{1}{l}{} & Original & \multicolumn{1}{l|}{} &                      & \multicolumn{3}{c}{BISM}                                           & \multicolumn{1}{c|}{} &                      & \multicolumn{3}{c}{DiSL}                                           &  \\
    & \multicolumn{1}{l|}{} &  & \multicolumn{1}{r}{}       & \multicolumn{1}{l|}{} & \multicolumn{1}{l}{} & \multicolumn{1}{r}{}      & \multicolumn{1}{l|}{} & \multicolumn{1}{l}{} & \multicolumn{1}{r}{}     & \multicolumn{1}{l|}{} & \multicolumn{1}{l}{} & KB       & \multicolumn{1}{l|}{} &                      & KB                   &                      & Ovh. \%               & \multicolumn{1}{c|}{} &                      & KB                   &                      & Ovh. \%                 &  \\ \hline
avrora  & \multicolumn{1}{l|}{} &  & 2.5 M                      & \multicolumn{1}{c|}{} &                      & 1550                      & \multicolumn{1}{c|}{} &                      & 35                       & \multicolumn{1}{c|}{} &                      & 257      & \multicolumn{1}{c|}{} &                      & 264                  &                      & 2.72                 & \multicolumn{1}{c|}{} &                      & 270                  &                      & 5.06                 &  \\
\rowcolor{gray!30}
batik   & \multicolumn{1}{l|}{} &  & 0.52 M                     & \multicolumn{1}{c|}{} &                      & 2689                      & \multicolumn{1}{c|}{} &                      & 136                      & \multicolumn{1}{c|}{} &                      & 1544     & \multicolumn{1}{c|}{} &                      & 1572                 &                      & 1.81                 & \multicolumn{1}{c|}{} &                      & 1588                 &                      & 2.85                 &  \\
fop     & \multicolumn{1}{l|}{} &  & 1.6 M                      & \multicolumn{1}{c|}{} &                      & 1336                      & \multicolumn{1}{c|}{} &                      & 172                      & \multicolumn{1}{c|}{} &                      & 1784     & \multicolumn{1}{c|}{} &                      & 1808                 &                      & 1.35                 & \multicolumn{1}{c|}{} &                      & 1876                 &                      & 5.16                 &  \\
\rowcolor{gray!30}
h2      & \multicolumn{1}{l|}{} &  & 28 M                       & \multicolumn{1}{c|}{} &                      & 472                       & \multicolumn{1}{c|}{} &                      & 61                       & \multicolumn{1}{c|}{} &                      & 694      & \multicolumn{1}{c|}{} &                      & 704                  &                      & 1.44                 & \multicolumn{1}{c|}{} &                      & 720                  &                      & 3.75                 &  \\
pmd     & \multicolumn{1}{l|}{} &  & 6.6 M                      & \multicolumn{1}{c|}{} &                      & 721                       & \multicolumn{1}{c|}{} &                      & 90                       & \multicolumn{1}{c|}{} &                      & 756      & \multicolumn{1}{c|}{} &                      & 774                  &                      & 2.38                 & \multicolumn{1}{c|}{} &                      & 794                  &                      & 5.03                 &  \\
\rowcolor{gray!30}
sunflow & \multicolumn{1}{l|}{} &  & 3.9 M                      & \multicolumn{1}{c|}{} &                      & 221                       & \multicolumn{1}{c|}{} &                      & 8                        & \multicolumn{1}{c|}{} &                      & 69       & \multicolumn{1}{c|}{} &                      & 71                   &                      & 2.90                 & \multicolumn{1}{c|}{} &                      & 74                   &                      & 7.25                 &  \\
xalan   & \multicolumn{1}{l|}{} &  & 1.04 M                     & \multicolumn{1}{c|}{} &                      & 661                       & \multicolumn{1}{c|}{} &                      & 9                        & \multicolumn{1}{c|}{} &                      & 100      & \multicolumn{1}{c|}{} &                      & 101                  &                      & 1.00                 & \multicolumn{1}{c|}{} &                      & 103                  &                      & 3.00                 &  \\
    &                       &  & \multicolumn{1}{r}{}       & \multicolumn{1}{l}{}  & \multicolumn{1}{l}{} & \multicolumn{1}{r}{}      & \multicolumn{1}{l}{}  & \multicolumn{1}{l}{} & \multicolumn{1}{r}{}     & \multicolumn{1}{l}{}  & \multicolumn{1}{l}{} &          & \multicolumn{1}{l}{}  & \multicolumn{1}{l}{} & \multicolumn{1}{l}{} & \multicolumn{1}{l}{} & \multicolumn{1}{l}{} & \multicolumn{1}{l}{}  & \multicolumn{1}{l}{} & \multicolumn{1}{l}{} & \multicolumn{1}{l}{} & \multicolumn{1}{l}{} & 
\end{tabular}
\end{table}
\egroup

Our evaluation confirms that BISM is a lightweight tool that can be used efficiently in runtime verification.
BISM incurs low overhead and produces fast and minimal bytecode.
%
% For the difference in bytecode size with DiSL, we observe that even with exception-handlers turned off in DiSL. 
% %
% DiSL still wraps a targeted region with try-finally blocks when using @After annotation, to guarantee that an event will be emitted 
%  after a method call, even if an exception is thrown.  
%
For the difference in bytecode size with DiSL, we observe that even with exception-handlers turned off, DiSL still wraps a targeted region with try-finally blocks when @After annotation is used. 
This is to guarantee that an event will be emitted after a method call, even if an exception is thrown.
For load-time instrumentation, the overhead gap closes between BISM and DiSL in benchmarks that have a large number of classes in scope and a small number of instrumented classes. 
That is because BISM performs a full analysis of the classes in scope to generate its static context. 
%
% While DiSL generates static context more efficiently by generating context only after marking the needed regions.
%
While DiSL generates static context only after marking the needed regions, which is more efficient.

\section{Related Work and Discussion} %{Comparison with Instrumentation Tools for Java} 
\label{sec:related-work}
We compare BISM with general-purpose tools for instrumenting Java programs. 

ASM~\cite{BrunetonASM02} is a bytecode manipulation framework utilized by several tools, including BISM.
ASM offers two APIs that can be used interchangeably to parse, load, and modify classes.
However, to use ASM, a developer has to deal with the low-level details of bytecode instructions and the JVM.
BISM offers extended ASM compatibility and provides abstraction with its aspect-oriented paradigm. 

DiSL is a bytecode-level instrumentation framework designed for dynamic program analysis~\cite{MarekVZABQ12}.
DiSL adopts an aspect-oriented paradigm. 
It provides an extensible joinpoint model and access to static and dynamic context information. 
Even though BISM provides a fixed set of joinpoints and static context objects, it performs static analysis on target programs 
to offer additional and needed out-of-the-box control-flow joinpoints with full static information.  
As for dynamic context objects, both BISM and DiSL provide equal access. However, DiSL provides typed dynamic objects. 
Also, both are capable of inserting synthetic local variables (restricted to primitive types in BISM). 
Both BISM and DiSL require basic knowledge about bytecode semantics from their users.
In DiSL, writing custom markers and context objects also requires additional ASM syntax knowledge.
However, DiSL does not allow the insertion of arbitrary bytecode instructions but provides a mechanism to write custom transformers in ASM that runs before instrumentation.  
Whereas, BISM allows to directly insert bytecode instructions, as such a mechanism is essential in many runtime monitoring scenarios, as seen in \secref{sec:aes}. 
All in all, DiSL provides more features (mostly targeted for writing dynamic analysis tools) and enables dynamic dispatch amongst multiple instrumentations and analysis without interference~\cite{BinderMTA16}, while BISM is more lightweight as shown by our evaluation.

AspectJ~\cite{KiczalesHHKPG01} is the standard aspect-oriented programming~\cite{KiczalesLMMLLI97} framework highly adop\-ted for instrumenting Java applications. 
It provides a high-level language used in several domains like monitoring, debugging, and logging.
AspectJ cannot capture bytecode instructions and basic blocks directly, forcing developers to insert additional code (like method calls) to the source program.
With BISM, developers can target single bytecode instructions and basic block levels, and also have access to local variables and stack values.
Furthermore, AspectJ introduces a significant instrumentation overhead and provides less control on where instrumentation snippets get inlined.
In BISM, the instrumentation methods are weaved with minimal bytecode instructions and are always inlined next to the targeted regions.

\section{Conclusion}
\label{sec:conclusion}
This paper introduces BISM (Bytecode-Level  Instrumentation for Software Monitoring), a lightweight bytecode instrumentation tool that features an expressive high-level instrumentation language inspired by the AOP paradigm. 
Overall, BISM is an effective tool for low-level and control-flow aware instrumentation, complementary to DiSL which is better suited for dynamic analysis (e.g. profiling).
We believe that BISM can be used for lightweight and expressive runtime verification. 
%They also show that BISM performs better than DiSL in general

%, although, in load-time mode, the overhead gap between them is small in benchmarks that have a high number of classes in scope and a low number of instrumented classes. 

% In the future, we plan to add more features to BISM. 
% %
% We also plan to use BISM to instrument some real applications for various purposes, such as logging information and verifying some security properties. 
% %
% 
\newpage
\bibliographystyle{unsrt}
\bibliography{biblio}

\begin{thebibliography}{10}

\bibitem{BartocciFFR18}
Ezio Bartocci, Yli{\`{e}}s Falcone, Adrian Francalanza, and Giles Reger.
\newblock Introduction to runtime verification.
\newblock In {\em Lectures on Runtime Verification - Introductory and Advanced
  Topics}. 2018.

\bibitem{KiczalesLMMLLI97}
Gregor Kiczales, John Lamping, Anurag Mendhekar, Chris Maeda, Cristina~Videira
  Lopes, Jean{-}Marc Loingtier, and John Irwin.
\newblock Aspect-oriented programming.
\newblock In Mehmet Aksit and Satoshi Matsuoka, editors, {\em ECOOP'97}, volume
  1241 of {\em LNCS}, pages 220--242. Springer, 1997.

\bibitem{FalconeKRT18}
Yli{\`{e}}s Falcone, Srdan Krstic, Giles Reger, and Dmitriy Traytel.
\newblock A taxonomy for classifying runtime verification tools.
\newblock In Christian Colombo and Martin Leucker, editors, {\em Runtime
  Verification - 18th International Conference, {RV} 2018, Limassol, Cyprus,
  November 10-13, 2018, Proceedings}, volume 11237 of {\em Lecture Notes in
  Computer Science}, pages 241--262. Springer, 2018.

\bibitem{BartocciFBCDHJK19}
Ezio Bartocci, Yli{\`{e}}s Falcone, Borzoo Bonakdarpour, Christian Colombo,
  Normann Decker, Klaus Havelund, Yogi Joshi, Felix Klaedtke, Reed Milewicz,
  Giles Reger, Grigore Rosu, Julien Signoles, Daniel Thoma, Eugen Zalinescu,
  and Yi~Zhang.
\newblock First international competition on runtime verification: rules,
  benchmarks, tools, and final results of {CRV} 2014.
\newblock {\em Int. J. Softw. Tools Technol. Transf.}, 21(1):31--70, 2019.

\bibitem{KiczalesHHKPG01}
Gregor Kiczales, Erik Hilsdale, Jim Hugunin, Mik Kersten, Jeffrey Palm, and
  William~G. Griswold.
\newblock Getting started with {AspectJ}.
\newblock {\em Commun. {ACM}}, 44(10):59--65, 2001.

\bibitem{BrunetonASM02}
Eric Bruneton, Romain Lenglet, and Thierry Coupaye.
\newblock {ASM}: A code manipulation tool to implement adaptable systems.
\newblock In {\em Adaptable and extensible component systems}, 2002.

\bibitem{BCEL}
Apache commons.
\newblock \url{https://commons.apache.org}.
\newblock Accessed: 2020-06-18.

\bibitem{MarekVZABQ12}
Luk{\'{a}}s Marek, Alex Villaz{\'{o}}n, Yudi Zheng, Danilo Ansaloni, Walter
  Binder, and Zhengwei Qi.
\newblock Disl: a domain-specific language for bytecode instrumentation.
\newblock In Robert Hirschfeld, {\'{E}}ric Tanter, Kevin~J. Sullivan, and
  Richard~P. Gabriel, editors, {\em Proceedings of the 11th International
  Conference on Aspect-oriented Software Development, {AOSD}, Potsdam,
  Germany}, pages 239--250. {ACM}, 2012.

\bibitem{DaCapo06}
Stephen~M. Blackburn, Robin Garner, Chris Hoffmann, Asjad~M. Khan, Kathryn~S.
  McKinley, Rotem Bentzur, Amer Diwan, Daniel Feinberg, Daniel Frampton,
  Samuel~Z. Guyer, Martin Hirzel, Antony~L. Hosking, Maria Jump, Han~Bok Lee,
  J.~Eliot~B. Moss, Aashish Phansalkar, Darko Stefanovic, Thomas VanDrunen,
  Daniel von Dincklage, and Ben Wiedermann.
\newblock The dacapo benchmarks: java benchmarking development and analysis.
\newblock In Peri~L. Tarr and William~R. Cook, editors, {\em Proceedings of the
  21th Annual {ACM} {SIGPLAN} Conference on Object-Oriented Programming,
  Systems, Languages, and Applications, {OOPSLA} 2006, October 22-26, 2006,
  Portland, Oregon, {USA}}, pages 169--190. {ACM}, 2006.

\bibitem{bism}
{BISM}: {Bytecode}-{Level} {Instrumentation} for {Software} {Monitoring}.

\bibitem{BinderMTA16}
Walter Binder, Philippe Moret, {\'{E}}ric Tanter, and Danilo Ansaloni.
\newblock Polymorphic bytecode instrumentation.
\newblock {\em Softw. Pract. Exp.}, 46(10):1351--1380, 2016.

\end{thebibliography}
\end{document}